\documentclass[Physsubmission, Phys]{SciPost}

\hypersetup{
    colorlinks,
    linkcolor={red!50!black},
    citecolor={blue!50!black},
    urlcolor={blue!80!black}
}

\usepackage[bitstream-charter]{mathdesign}
\DeclareSymbolFont{usualmathcal}{OMS}{cmsy}{m}{n}
\DeclareSymbolFontAlphabet{\mathcal}{usualmathcal}

\begin{document}

\begin{center}{\Large \textbf{
Light meson decays at BESIII\\
}}\end{center}

\begin{center}
Xiao Lin Kang\textsuperscript{1$\star$}\\
on behalf of BESIII Collaboration
\end{center}

\begin{center}
{\bf 1} China University of GeoSciences, 388 Lumo Rd, Wuhan, P. R. China
\\
* kangxiaolin@cug.edu.cn
\end{center}

\begin{center}
\today
\end{center}


\definecolor{palegray}{gray}{0.95}
\begin{center}
\colorbox{palegray}{
  \begin{tabular}{rr}
  \begin{minipage}{0.1\textwidth}
    \includegraphics[width=30mm]{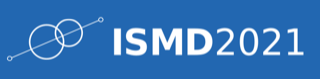}
  \end{minipage}
  &
  \begin{minipage}{0.75\textwidth}
    \begin{center}
    {\it 50th International Symposium on Multiparticle Dynamics}\\ {\it (ISMD2021)}\\
    {\it 12-16 July 2021} \\
    \doi{10.21468/SciPostPhysProc.?}\\
    \end{center}
  \end{minipage}
\end{tabular}
}
\end{center}

\section*{Abstract}
{\bf
The world's largest sample of $J/\psi$ events accumulated at the BESIII dector offers a
unique opportunity to investigate $\eta$ and $\eta^\prime$ physics via two body $J/\psi$
radiative or hadronic decays. In recent years the BESIII experiment has made significant
progresses in $\eta/\eta^\prime$ decays. A selection of recent highlights in light meson
spectroscopy at BESIII are reviewed in this report, including the observation of
$\eta^\prime\rightarrow\pi^+\pi^-\mu^+\mu^-$, the study of $\eta^\prime\rightarrow\pi^+\pi^-e^+e^-$.
and search for $CP$-violation in this decay, search for the rare decays of
$\eta^\prime\rightarrow4\pi^0$ and $\eta^\prime\rightarrow\gamma\gamma\eta$, as well as
the precision measurement of the branching fraction of $\eta^\prime$ decays.
}


\noindent\rule{\textwidth}{1pt}

\section{Introduction}\label{sec:intro}

Because of the special role in understanding low energy quantum chromodynamics (QCD),
$\eta$ and $\eta^\prime$ mesons attract considerable theoretical and experimental attention.
As a mixture of the lowest pseudoscale singlet and octet, $\eta/\eta^\prime$ have inspired
a wide variety of physics issues, e.g., $\eta-\eta^\prime$ mixing, the light quark masses,
the fundamental discrete symmetries, as well as  physics beyond the standard model (SM).
In addition, $\eta/\eta^\prime$ decays offer unique opportunities to investigate decay
dynamics and test different chiral perturbation theory (ChPT) and the vector meson
dominance (VMD) models. Moreover, it is also possible to search for new phenomena
in rare or forbidden $\eta/\eta^\prime$ decays.

The BESIII detector~\cite{BESIII:2009fln} is operated at BEPCII, an $e^+e^-$ collider running at a center of
mass energy of $2-4.9$ GeV. BESIII experiment accumulated 1.31 billion $J/\psi$ events
in the years 2009 and 2012. The results reported in this proceeding are based on this
data sample. In 2018 and 2019, BESIII continue collected data at $J/\psi$ events, makes
the number of $J/\psi$ increased to 10 billion in total. 
Copious $\eta$ and $\eta^\prime$ mesons are produced via the radiative and hadronic
decays of $J/\psi$. Considering the radiative decays of $J/\psi$,
$J/\psi\rightarrow\gamma\eta/\eta^\prime$, the total $J/\psi$ sample corresponds to
$1.1\times10^7$ $\eta$ mesons and $5.2\times10^7$ $\eta^\prime$ mesons, respectively,
which allow to analysis specific decays with the unprecedented statistics and to
search for rare and forbidden decays.

\section{Observation of $\eta^\prime\rightarrow\pi^+\pi^-\mu^+\mu^-$} \label{sec:sect_ppmm}

The decays $\eta^\prime\rightarrow\pi^+\pi^-l^+l^-$ (with $l=e$ or $\mu$) are
expeected to have similar structure as $\eta^\prime\rightarrow\pi^+\pi^-\gamma$ with the
radiative $\gamma$ replaced with an offshell one and decays into a lepton pair~\cite{Witten:1983tw}.
These decays may involve the box anomaly contribution and have been investigated theoretically
with different models, including the effective meson theory, the chiral unitary approach, and
the hidden gauge model, {\it et al}. Due to the lower phase space of
$\eta^\prime\rightarrow\pi^+\pi^-\mu^+\mu^-$, the branching fraction is predicted  in the range
of $(1.5-2.5)\times10^{-5}$~\cite{Faessler:1999de,Borasoy:2007dw,Petri:2010ea}, which are about
2 orders of magnitude lower than those for $\eta^\prime\rightarrow\pi^+\pi^-e^+e^-$.

\begin{figure}[h]
\centering
\includegraphics[width=0.43\textwidth]{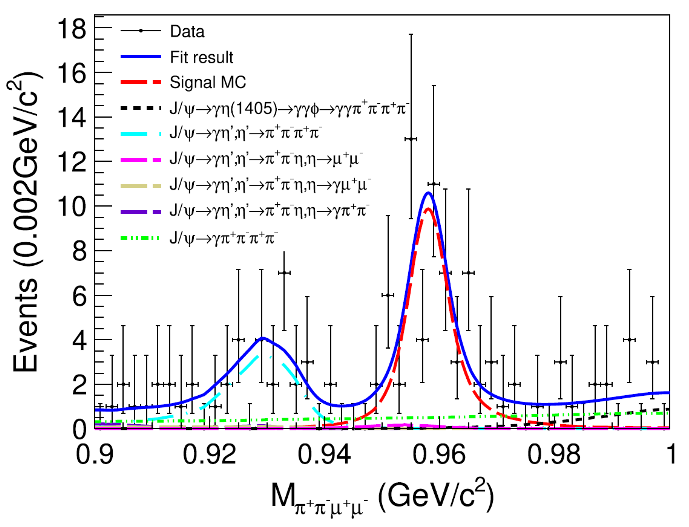}
\caption{The $\pi^+\pi^-\mu^+\mu^-$ invariant mass spectrum around the $\eta^\prime$ mass.
The dots with error bars represent the data, the red line is signal MC, and the blue line
is the total fit result. The other dotted lines represent different backgrounds.
The plot is reproduced from Ref~\cite{BESIII:2020elh} under the Creative Commons Attribution 4.0 International license.}
\label{fig:Mppmm}
\end{figure}

BESIII has performed a search for this decay with a data sample of 225 million $J/\psi$
events and set the upper limit ${\cal B}(\eta^\prime\rightarrow\pi^+\pi^-\mu^+\mu^-)<2.9
\times10^{-5}$~\cite{BESIII:2013tjj} at 90\% confidence level (CL), which is the most stringent result to data.
Using a sample of $1.31\times10^9$ $J/\psi$ events, BESIII updated this analysis~\cite{BESIII:2020elh}
recently via the process $J/\psi\rightarrow\gamma\eta^\prime$. Clear $\eta^\prime$ signal with a
significance of $8\sigma$ is observed for the first time in the invariant mass of
$\pi^+\pi^-\mu^+\mu^-$, as the right peak in Fig.~\ref{fig:Mppmm} shown. 
A global fit to the $\pi^+\pi^-\mu^+\mu^-$ mass spectrum yields $53\pm9$ signal events. 
The corresponding branching fraction is determined to be $(1.97\pm0.33\pm0.19)\times10^{-5}$,
where the first uncertainty is statistical and the second systematical.
The result is in good agreement with theoretical predictions.

\section{The study of $\eta^\prime\rightarrow\pi^+\pi^-e^+e^-$} \label{sec:sect_ppee}

As described in Sec.~\ref{sec:sect_ppmm}, this process is expected to proceed through
an intermediate virtual photon $\eta^\prime\rightarrow\pi^+\pi^-\gamma^*\rightarrow
\pi^+\pi^-e^+e^-$. By comparing the precision measurement of the branching ratio of
$\eta^\prime\rightarrow\pi^+\pi^-e^+e^-$ with the predictions from different theoretical
approaches, such as VMD and ChPT models, it is possible to probe the
electromagnetic structure of $\eta^\prime$ meson.

\begin{figure}[h]
\centering
\includegraphics[width=0.43\textwidth]{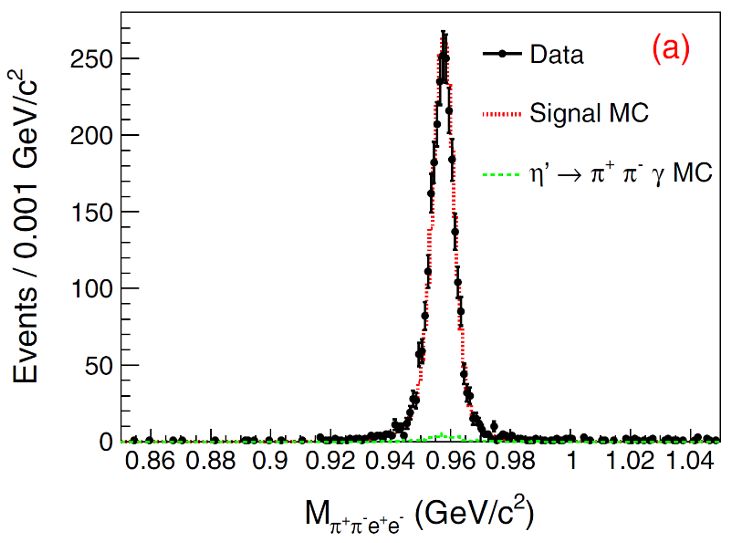}
\includegraphics[width=0.43\textwidth]{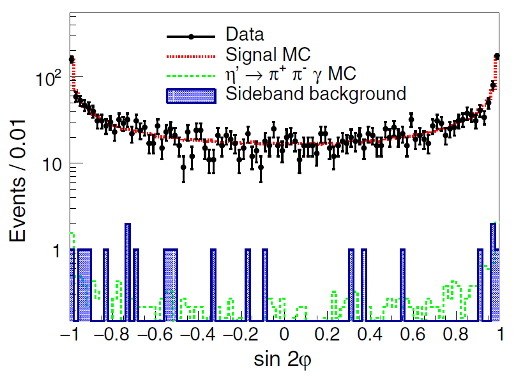}
\caption{The $\pi^+\pi^-e^+e^-$ invariant mass spectrum around $\eta^\prime$ mass (left) and
the $\sin2\phi$ distribution in the signal region (right). The dots with error bars are data,
the red histograms are signal MC, the green histograms are backgrounds from $\eta^\prime\rightarrow
\pi^+\pi^-\gamma$, and the blue histogram in right panel is background from $\eta^\prime$ sidebands.
The plot is reproduced from Ref~\cite{BESIII:2020otu} under the Creative Commons Attribution 4.0 International license.}
\label{fig:Mppee}
\end{figure}

In 2013, BESIII studied this decay with a data sample of 225 million $J/\psi$ events
and the branching fraction is determined to be $(2.11\pm0.12(stat)\pm0.14(syst))\times10^{-3}$~\cite{BESIII:2013tjj}.
Using a data sample of 1.31 billion $J/\psi$ events, BESIII updated this analysis recently~\cite{BESIII:2020otu}
via the process $J/\psi\rightarrow\gamma\eta^\prime$. Clean $\eta^\prime$ signal is observed
in the invariant mass of $\pi^+\pi^-e^+e^-$, as shown in the left panel of Fig.~\ref{fig:Mppee}.
The background contamination ratio is around 2\%, mainly from $\eta^\prime\rightarrow\pi^+\pi^-\gamma$.
A global fit to the $\pi^+\pi^-e^+e^-$ mass spectrum yields $2584\pm52$ signal events,
corresponds to ${\cal B}(\eta^\prime\rightarrow\pi^+\pi^-e^+e^-)=(2.42\pm0.05(stat)\pm0.08(syst))\times10^{-3}$.
The statistical uncertainty has been improved by a factor of two compared to the previous BESIII result,
which is superseded by this result.
Reference~\cite{Petri:2010ea} predicts the branching fraction based on two different VMD models,
$(2.17\pm0.21)\times10^{-3}$ from the hidden gauge model and $(2.27\pm0.13)\times10^{-3}$ from
modified VMD model, respetively.
While the unitary chiral perturbation theory approach yields a branching fraction of $(2.13^{+0.17}_{-0.31})\times10^{-3}$~\cite{Borasoy:2007dw}.
Our result is consistent with all three predictions, and about one standard deviation higher than each of them.

In addition, a possible $CP$-violating mechanism~\cite{Gao:2002gq} has been proposed for
this decay, which is beyond the most widely studied flavor changing neutral process. The
decay can be proceed via the parity-conserving magnetic amplitudes and the parity-violating
electric amplitudes. $CP$-violation effect could be induced from the interference among
the two amplitudes and tested by an asymmetry in the angle $\phi$ between the pions and
electrons decay planes in the $\eta^\prime$ rest frame, defined as
\begin{equation}
A_\phi  = \frac{ N(\sin2\phi>0) - N(\sin2\phi<0) }{ N(\sin2\phi>0) + N(\sin2\phi<0) },
\end{equation}
where $N(x)$ is the acceptance-corrected number of events in the corresponding angular region.
The decay plane asymmetry $A_\phi$ has been evaluated for the events in the signal region
after background subtraction. The obtained value is $A_\phi=(2.9\pm3.7_{stat}\pm1.1_{syst})\%$,
which is extracted for the first time and consistent with 0 within uncertainties. The distribution
of the $\sin2\phi$ variable is shown
in the right panel of Fig~\ref{fig:Mppee}.

\section{The rare decay of $\eta^\prime\rightarrow\pi^0\pi^0\pi^0\pi^0$} \label{sec:sect_4pi0}

In theory the decay $\eta^\prime\rightarrow\pi^0\pi^0\pi^0\pi^0\pi^0$ is highly suppressed because of
the $S$-wave $CP$-violation. In the effective chiral Lagrangian approach,
the $S$-wave in $\eta^\prime\rightarrow\pi^0\pi^0\pi^0\pi^0$ is induced by the so-called
$\theta$-term in the QCD Lagrangian to account for the solution of the strong-$CP$ problem,
with an expected branching ratio at the level of $10^{-23}$~\cite{Pich:1991fq,Ottnad:2009jw}.
While higher-order contributions involving a $D$-wave pion-pion charge-exchange rescattering
loop provide a $CP$-conserving route, and the branching ratio is calculated around
$\sim4\times10^{-8}$~\cite{Guo:2011ir}.
An alternative mechanism through two $f_2$ tensor mesons is found to be completely negligible
in comparison~\cite{Guo:2011ir}.
However, due to the lack of knowledge at such a high order in the chiral expansion and
the use of a model to make an estimation, the theoretical prediction is not strinctly based
on the effective field theory. A search for the decay $\eta^\prime\rightarrow\pi^0\pi^0\pi^0\pi^0$
is useful to check the reliability.
The most stringent upper limit $3.2\times10^{-4}$ at 90\% CL was from
the GAMS-4$\pi$ Collaboration~\cite{Donskov:2014zja}.

\begin{figure}[h]
\centering
\includegraphics[width=0.43\textwidth]{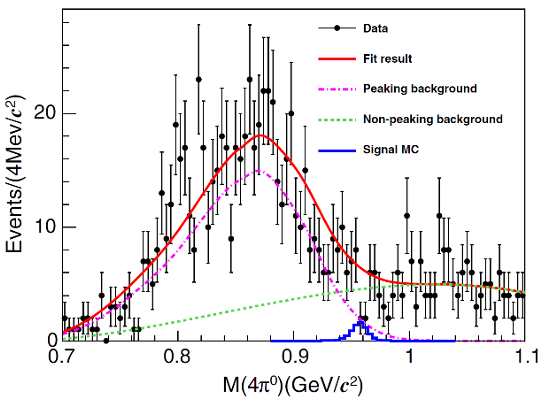}
\caption{The $\pi^0\pi^0\pi^0\pi^0$ invariant mass spectrum in data (black dots with error bars),
together with the total fit result (red line) and the contributions from
different backgrounds (green and pink lines).
The expected shape of the signal
contribution is also shown as the blue histogram with arbitrary normalization.
The plot is reproduced from Ref~\cite{BESIII:2019tel} under the Creative Commons Attribution 4.0 International license.}
\label{fig:M4pi0}
\end{figure}

Using a data sample of $1.31\times10^9$ $J/\psi$ events, BESIII Collaboration performed
the search for the rare decay $\eta^\prime\rightarrow4\pi^0$ via $J/\psi\rightarrow
\gamma\eta^\prime$~\cite{BESIII:2019tel}. No significant $\eta^\prime$ signal is observed in the $4\pi^0$
invariant mass spectrum, shown in Fig.~\ref{fig:M4pi0}. With a Bayesian approach, the
upper limit on the branching
fraction is determined to be ${\cal B}(\eta^\prime\rightarrow4\pi^0)<4.94\times10^{-5}$
at 90\% CL, which is a factor of 6 smaller than the previous experimental limit.
The current limit is still far to reach the theoretical predication with a level
of $10^{-8}$. Further studies of $\eta^\prime$ rare decays are still necessary
to test the ChPT and VMD model and look for the $CP$-violation (S-wave)
$\eta^\prime\rightarrow4\pi^0$ decay.

\section{Search for the rare decay of $\eta^\prime\rightarrow\gamma\gamma\eta$} \label{sec:sect_ggeta}

Within the frameworks of the linear $\sigma$ model and
VMD model, the branching fractions of $\eta^\prime\rightarrow\gamma\gamma\pi^0$
and $\eta^\prime\rightarrow\gamma\gamma\eta$ are predicted to be $2.91(21)\times10^{-3}$
and $1.17(8)\times10^{-4}$, respectively~\cite{Escribano:2018cwg}.
The branching fraction of $\eta^\prime\rightarrow\gamma\gamma\pi^0$ was determined to be $(32.0\pm0.7\pm2.3)\times10^{-4}$~\cite{BESIII:2016oet} by BESIII experiment, while the $\eta^\prime\rightarrow\gamma\gamma\eta$ decay has not been observed to date.
The most stringent upper limit $8\times10^{-4}$ at the 90\%
CL was from GAMS-4$\pi$ Collaboration~\cite{Donskov:2015epm}.

Using a data sample of $1.31\times10^9$ $J/\psi$ events, a search for the rare
decay $\eta^\prime\rightarrow\gamma\gamma\eta$ is performed by BESIII experiment
via $J/\psi\rightarrow\gamma\eta^\prime$~\cite{BESIII:2019ofm}. A global fit to the $\gamma\gamma\eta$
invariant mass spectrum yields $24.9\pm10.3$ $\eta^\prime\rightarrow\gamma\gamma\eta$ events,
with a statistical significance of $2.6\sigma$, and the branching fraction is calculated to be
$(8.25\pm3.41\pm0.72)\times10^{-5}$, which need to be confirmed with higher statistics.
Here the first error is statistical and the second systematical. An upper limit of the branching
fraction is also set as $1.3\times10^{-4}$ at 90\% CL, which is consistent with a recent theoretical
prediction of $2\times10^{-4}$~\cite{Escribano:2018cwg} within the frame work of the linear 
$\sigma$ model and the VMD model.

\begin{figure}[h]
\centering
\includegraphics[width=0.43\textwidth]{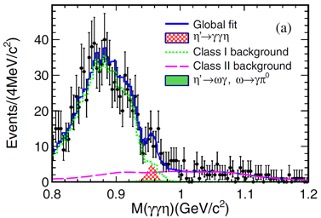}
\caption{Results of the fit to $\gamma\gamma\eta$ invariant mass distribution. The black dots with error bars are data,
and the others are backgrounds.
The plot is reproduced from Ref~\cite{BESIII:2019ofm} under the Creative Commons Attribution 4.0 International license.}
\label{fig:Mggeta}
\end{figure}

\section{Precision measurement of the branching fractions of $\eta^\prime$ decays} \label{sec:sect_bretap}

Due to the difficulty of tagging $\eta^\prime$ inclusive decays, the absolute branching
fractions (BFs) of its decays have yet been measured. The exclusive BFs of the $\eta^\prime$
summarized by the Particle Data Group (PDG)~\cite{PDG:2020ssz} are all relative measurements.
The absolute BF measurement of the five dominant decay modes are also essential in order
to improve the precision of the BFs for seveal $\eta^\prime$ decays, which are obtained
via normalization to the dominant $\eta^\prime$ decay modes.

Using a data sample of $1.31\times10^9$ $J/\psi$ events, the BESIII Collaboration develops
an approach to measure the absolute BFs of the exclusive decays of the $\eta^\prime$ meson~\cite{BESIII:2019gef}.
A model-independent measurement of the BF for $J/\psi\rightarrow\gamma\eta^\prime$ is performed
by analyzing events where the radiative photon converts into an $e^+e^-$ pair. The BF of 
$J/\psi\rightarrow\gamma\eta^\prime$ is determined to be $(5.27\pm0.03\pm0.05)\times10^{-3}$,
which is in agreement with the world average value, with a significantly improved precision.
After the $\eta^\prime$ inclusive measurement, $\eta^\prime$ decays to $\gamma\pi^+\pi^-$,
$\eta\pi^+\pi^-$, $\eta\pi^0\pi^0$, $\gamma\omega$, and $\gamma\gamma$ candidates are selected
via $J/\psi\rightarrow\gamma\eta^\prime$ with the radiative photon detected by the
eclectromagnetic calorimeter.
Together with the $\eta^\prime$ sample tagged by photon conversion, the absolute BFs of five
dominant decays of the $\eta^\prime$ are determined and presented in Table~\ref{tab:BFetap},
which is the first independent measurements. In addition, relative BFs for $\eta^\prime$ decays
are also presented in Table~\ref{tab:BFetap}, which are in agreement w.r.t. CLEO's result~\cite{CLEO:2009cot}
within two standard deviation but with improved precision.
 
\begin{table}[h]
\centering
 \caption{\label{fitres} Summary of the measured BFs for $\eta^\prime$ decays.}
 \resizebox{\textwidth}{!}{%
 \begin{tabular}{l|cc|cc}\hline\hline
    & \multicolumn{2}{|c|}{${\cal B}(\eta^\prime\rightarrow X)(\%)$} & \multicolumn{2}{|c}{${\cal B}/{\cal B}(\eta^\prime\rightarrow\eta\pi^+\pi^-)$}    \\ \cline{2-5}
  Decay mode &  BESIII~\cite{BESIII:2019gef} & PDG~\cite{PDG:2020ssz}  & BESIII~\cite{BESIII:2019gef}  & CLEO~\cite{CLEO:2009cot} \\\hline

 $\eta^\prime\rightarrow\gamma\pi^+\pi^-$& $29.90\pm0.03\pm0.55$  & $28.9\pm0.5$ & $0.725\pm0.002\pm0.010$   & $0.677\pm0.024\pm0.011$\\
 $\eta^\prime\rightarrow\eta\pi^+\pi^-$  & $41.24\pm0.08\pm1.24$  & $42.6\pm0.7$ & - & - \\
 $\eta^\prime\rightarrow\eta\pi^0\pi^0$  & $21.36\pm0.10\pm0.92$  & $22.8\pm0.8$ & $0.518\pm0.003\pm0.021$   & $0.555\pm0.043\pm0.013$\\
 $\eta^\prime\rightarrow\gamma\omega$    & $2.489\pm0.018\pm0.074$& $2.62\pm0.13$& $0.0604\pm0.0005\pm0.0012$& $0.055\pm0.007\pm0.001$\\
 $\eta^\prime\rightarrow\gamma\gamma$    & $2.331\pm0.012\pm0.035$& $2.22\pm0.08$& $0.0565\pm0.0003\pm0.0015$& $0.053\pm0.004\pm0.001$\\
  \hline\hline
	\end{tabular}
	}%
\label{tab:BFetap}
\end{table}

\section{Conclusion}

The BESIII collaboration has produced fruitful results related with light meson decays,
including the studies of the decay dynamics, tests of discrete symmetries, searches for
rare decays, and many other interesting results not covered in this proceeding. The
BESIII experiment has accumulated 10 billion $J/\psi$ events in total, which is a unique
worldwide sample, allows to study the light mesons with unprecedented statistics. Ongoing
analyses will produce more precise results in the next years.

%



\nolinenumbers

\end{document}